\documentclass[amsmath,longbibliography,superscriptaddress,reprint]{revtex4-2}

\usepackage{hyperref}
\usepackage{graphicx}
\usepackage{siunitx}

\begin{document}

\title{Intrinsic process for upconversion photoluminescence via $K$-momentum phonon coupling in carbon nanotubes}

\author{Daichi Kozawa}
\email[Corresponding Author: ]{kozawa.daichi@nims.go.jp}
\affiliation{Quantum Optoelectronics Research Team,
    RIKEN Center for Advanced Photonics, Wako, Saitama 351-0198, Japan}
\affiliation{Nanoscale Quantum Photonics Laboratory,
    RIKEN Cluster for Pioneering Research, Wako, Saitama 351-0198, Japan}
\affiliation{Research Center for Materials Nanoarchitectonics, National Institute for Materials Science, Tsukuba, Ibaraki 305-0044, Japan}

\author{Shun Fujii}
\affiliation{Quantum Optoelectronics Research Team,
    RIKEN Center for Advanced Photonics, Wako, Saitama 351-0198, Japan}
\affiliation{Department of Physics, Faculty of Science and Technology, Keio University, Yokohama, Kanagawa 223-8522, Japan}

\author{Yuichiro K. Kato}
\email[Corresponding Author: ]{yuichiro.kato@riken.jp}
\affiliation{Quantum Optoelectronics Research Team,
    RIKEN Center for Advanced Photonics, Wako, Saitama 351-0198, Japan}
\affiliation{Nanoscale Quantum Photonics Laboratory,
    RIKEN Cluster for Pioneering Research, Wako, Saitama 351-0198, Japan}

\begin{abstract}
    We investigate the intrinsic microscopic mechanism of photon upconversion in air-suspended single-walled carbon nanotubes through photoluminescence and upconversion photoluminescence spectroscopy. Nearly linear excitation power dependence of upconversion photoluminescence intensity is observed, indicating a one-photon process as the underlying mechanism. In addition, we find a strongly anisotropic response to the excitation polarization which reflects the intrinsic nature of the upconversion process. In upconversion photoluminescence excitation spectra, three peaks are observed which are similar to photoluminescence sidebands of the $K$-momentum dark singlet exciton. The features in the upconversion photoluminescence excitation spectra are well reproduced by our second-order exciton-phonon interaction model, enabling the determination of phonon energies and relative amplitudes. The analysis reveals that the upconversion photoluminescence can be described as a reverse process of the sideband emission linked to the $K$-momentum phonon modes. The validity of our model is further reinforced by temperature-dependent upconversion photoluminescence excitation measurements reflecting variations in the phonon population. Our findings underscore the pivotal role of the resonant exciton-phonon coupling in pristine carbon nanotubes and presents potential for advanced optothermal technologies by engineering the excitation pathways.

\end{abstract}

\maketitle

\section{Introduction}
Upconversion photoluminescence (UCPL) is a nonlinear optical process in which photons with lower energy are absorbed and re-emitted as photons with higher energy~\cite{Eshlaghi2008, Jones2016, Manca2017, Jadczak2021,Akizuki2015a}. UCPL has profound implications in various fields, prominently in telecommunications~\cite{Chen2011}, advanced photonics~\cite{Gao2020}, renewable energy technologies~\cite{Ghazy2021}, and optical cooling~\cite{Epstein1995}. The unique ability of UCPL to manipulate photon energies not only contributes to the theoretical understanding of optical processes in materials but also paves the way for practical applications, such as enhancing solar cell efficiency~\cite{Ghazy2021} and developing new bio-imaging techniques~\cite{Aota2016b}. The capability of UCPL in bridging the gap between available and usable light spectra especially in the near-infrared range can facilitate these applications.

Near-infrared upconversion in single-walled carbon nanotubes (SWNTs) stands out for its high efficiency and remarkable energy gain exceeding several times the thermal energy at room temperature~\cite{Akizuki2015a, Aota2016b}. Previous research has focused on liquid-dispersed SWNTs with the limited chiralities of (6,5) and (8,3) where UCPL is attributed to one-phonon-assisted upconversion processes under one-photon excitation conditions~\cite{Akizuki2015a}. The efficiency of UCPL is found to be enhanced by localized states of intentionally introduced defects, suggesting that the UCPL is predominantly extrinsic. Pristine SWNTs without such defects thus are presumed to exhibit negligible UCPL.

Here we investigate UCPL in as-grown, air-suspended SWNTs which can be considered defect free except for the tube ends~\cite{Moritsubo2010,Ishii2015}. By performing UCPL and photoluminescence (PL) measurements, the spectra and the images confirm that the emission originates from the identical nanotube. The dependence of the UCPL intensity on the excitation power reflects the one-photon excitation process, and the dependence on the excitation polarization shows evidence for the intrinsic characteristics. UCPL spectra across various chiralities of SWNTs are also examined, revealing that UCPL is a universal phenomenon across the chiralities. Upconversion photoluminescence excitation (UCPLE) spectra are compared with sidebands in PL spectra to elucidate the excitation process in UCPL. We develop a theoretical model for upconversion that involves photon-exciton and exciton-phonon interactions, which is able to quantitatively explain the features in the UCPLE spectra. We conduct temperature-dependent UCPLE spectroscopy, verifying the model based on phonon-assisted upconversion in pristine SWNTs.

\section{Photoluminescence and upconversion photoluminescence measurements}
The air-suspended SWNTs are grown over trenches on Si substrates by chemical vapor deposition ~\cite{Ishii2015,Ishii2017,Ishii2019}. We fabricate the trenches with a depth of $\sim$1~$\mu$m and a width from 0.5 to 4.0~$\mu$m by performing electron-beam lithography and dry etching. Another electron-beam lithography step is conducted to define catalyst areas near the trenches, onto which Co- or Fe-silica catalyst dispersed in ethanol are spin-coated and then lifted off. SWNTs are synthesized over the trenches using alcohol chemical vapor deposition under a flow of ethanol with a carrier gas of Ar/H$_2$ at 800$^\circ$C for one minute.

A home-built confocal microscopy system is used to collect PL spectra in dry N$_2$ gas~\cite{Ishii2015}. Si substrates are mounted on a motorized three-dimensional translation stage, allowing us to perform automated measurements over hundreds of SWNTs. We utilize a continuous-wave Ti:sapphire laser for excitation and a liquid-N$_2$-cooled InGaAs photodiode array attached to a 30-cm spectrometer for detection. Laser polarization is kept perpendicular to the trenches unless specified otherwise. The excitation beam is focused via an objective lens with a numerical aperture of 0.65 and a focal length of 1.8~mm. The focused spot exhibits 1/$e^2$ diameters of 1.46, 1.31, and 1.06~$\mu$m for energies of 1.36, 1.46, and 1.60~eV, respectively, where these diameters are determined by PL line scans perpendicular to a suspended tube. The excitation polarization is rotated by a half-wave plate mounted on a motorized rotation stage for polarization-dependent PL measurements. All PL spectra are taken at the center of the nanotubes except for hyperspectral PL imaging.

UCPL spectra are also taken with the same microscope using a continuous-wave laser diode with an energy of 0.800~eV where the $1/e^2$ diameter of the focused beam is 2.61~$\mu$m. A short-pass dichroic filter with a transmission band between 0.855 and 1.24~eV is utilized to block the excitation light during UCPL collection. The excitation power is tuned by a fiber optic variable optical attenuator, and the excitation polarization is rotated by the half-wave plate.

\section{Upconversion photoluminescence in pristine nanotubes}

\begin{figure*}[t]
    \includegraphics{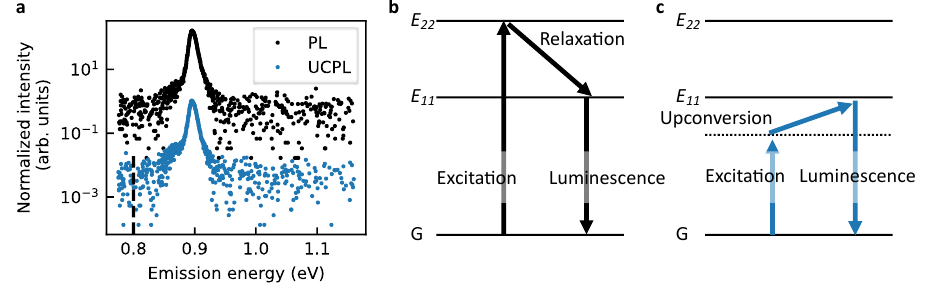}
    \caption{
        \label{fig1}
        (a) PL and UCPL spectra for the same (9,8) SWNT with intensities normalized by the excitation power density. The PL spectrum is collected with an excitation energy of 1.54~eV and a power of 10.0~$\mu$W. The UCPL spectrum is obtained with an excitation energy of 0.800~eV marked by a broken line and a power of 1000~$\mu$W.
        Energy diagrams for (b) the PL and (c) the UCPL processes. Solid horizontal lines represent the ground (G), the first sub-band ($E_{11}$), and the second sub-band exciton states ($E_{22}$), while the dotted line is an intermediate state in the upconversion process.
    }
\end{figure*}

We conduct PL and UCPL measurements on the same as-grown (9,8) SWNT where the chirality is determined by photoluminescence excitation (PLE) spectroscopy (Fig.~S1 in Supplemental Material~\cite{supp}). In Fig.~\ref{fig1}a, both spectra are obtained with an excitation polarization aligned to the nanotube axis and are displayed on a logarithmic scale. We observe UCPL in the pristine nanotube in contradiction to the expectation that UCPL is negligible in the absence of localized states ~\cite{Akizuki2015a}. Here the PL and UCPL spectra are compared to confirm that the UCPL arises from $E_{11}$ exciton emission. The emission peak mirrors the $E_{11}$ PL emission in terms of energy and linewidth. The UCPL intensity is lower than the PL intensity by a factor of 155 but with a significant energy gain of 96.2~meV relative to the excitation energy.

The observed PL and UCPL processes are illustrated by energy diagrams in Fig.~\ref{fig1}b and~\ref{fig1}c. The PL process begins with excitation of an exciton to an energy of $E_{22}$ and the exciton then relaxes to the $E_{11}$ state, recombining to emit a photon with an energy of $E_{11}$. The UCPL involves excitation to an intermediate state and subsequent upconversion to the $E_{11}$ state before luminescence. We note that UCPL via real states is enhanced by chemically introducing defects into solution-processed SWNTs~\cite{Akizuki2015a}, but UCPL in pristine SWNTs is considerably strong. The intermediate state in air-suspended SWNTs is likely to be virtual, given that the effect of defects is minimal in as-grown nanotubes~\cite{Ishii2015}.

\begin{figure}[t]
    \includegraphics{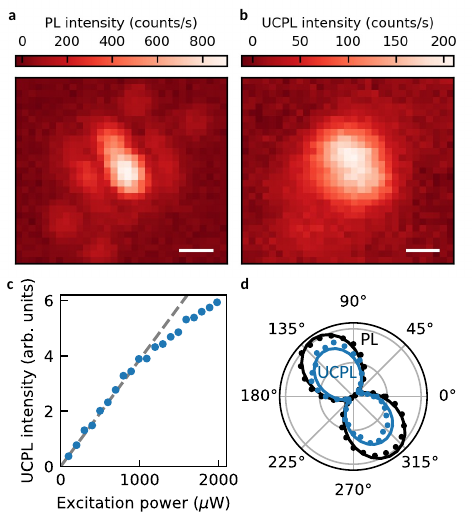}
    \caption{
        \label{fig2}
        (a) PL and (b) UCPL images of the identical (9,8) SWNT suspended over 3-$\mu$m trench, in which the images are obtained with the spectrally integrated intensity within a 3-meV window at the $E_{11}$ emission peak. For PL, an excitation energy of 1.54~eV and a power of 10.0~$\mu$W are used, while for UCPL, an excitation energy of 0.800~eV is applied, with a power of 1000~$\mu$W. Scale bars represent 1~$\mu$m.
        (c) The excitation power dependence of UCPL intensity with an excitation energy of 0.800~eV. The broken line is a fit up to the power of 1090~$\mu$W.
        (d) Polarization dependence of PL and UCPL intensity. Solid lines are fits to the expression described  main text. PL and UCPL intensities in panels (c, d) are derived from the peak areas calculated via a Lorentzian fit to the emission spectra.
    }
\end{figure}

Excitation imaging measurements are performed to characterize the spatial distributions of $E_{11}$ emission for PL and UCPL. We scan over the same (9,8) SWNT to collect emission and construct intensity maps (Fig.~\ref{fig2}a and \ref{fig2}b). The spatial overlap of the bright regions in the intensity maps confirm that the origin of the emission is the same nanotube. Both images visualize the long SWNT suspended over a trench, consistent with the picture that PL brightness is influenced by exciton diffusion and end quenching~\cite{Moritsubo2010,Ishii2015,Xie2012,Anderson2013}. The UCPL image has a lower resolution because the focused laser spot for UCPL measurements is larger by a factor of $\sim$2 than for PL.

Well-known mechanisms of upconversion involve one-photon and two-photon absorption processes characterized by linear and quadratic response, respectively, in the UCPL intensity to the excitation power~\cite{Zhang2019}. We collect UCPL spectra with various powers up to $\sim$2000~$\mu$W (Fig.~S2 in Supplemental Material~\cite{supp}) and extract the peak area using a Lorentzian fit to the $E_{11}$ emission peak (Fig.~\ref{fig2}c). Our observation of the mostly linear response can be explained by the one-photon process facilitated by exciton-phonon scattering. This finding is in contrast with the expected quadratic response in two-photon excitation and Auger recombination-mediated upconversion~\cite{Ma2005}. A subtle shift towards a sublinear trend is observed at powers above 1000~$\mu$W, indicating the onset of the exciton-exciton annihilation regime~\cite{Ishii2015, Moritsubo2010}.

Excitation polarization dependence of the UCPL intensity can be an evidence for intrinsic characteristics of the upconversion process. Figure~\ref{fig2}d shows the excitation polarization dependence of PL and UCPL intensities for the same (9,8) SWNT. We fit the data to $I_\mathrm{min}+(I_\mathrm{max}-I_\mathrm{min}) \sin^2(\phi + \phi_0)$, where $I_\mathrm{min}$ and $I_\mathrm{max}$ correspond to the minimum and maximum intensity, respectively, $\phi$ is the excitation polarization angle, and $\phi _0$ is the suspended angle of the SWNT with respect to the length of the trench. From the fit parameters, we compute the degree of polarization $(I_\mathrm{max}-I_\mathrm{min})/(I_\mathrm{max}+I_\mathrm{min})$ to be 0.82$\pm$0.11 for PL and 0.77$\pm$0.12 for UCPL. The anisotropic response reflects the one-dimensional character of SWNTs~\cite{Lefebvre2004, Bozovic2000, Ajiki1993}, supporting the intrinsic nature of UCPL in air-suspended SWNTs. We note that the polarization degree of UCPL in solution-processed SWNTs is lower than that of PL~\cite{Aota2016b}, which is attributed to the extrinsic nature of UCPL mediated by defect states.

\begin{figure}
    \includegraphics[width=7.6cm]{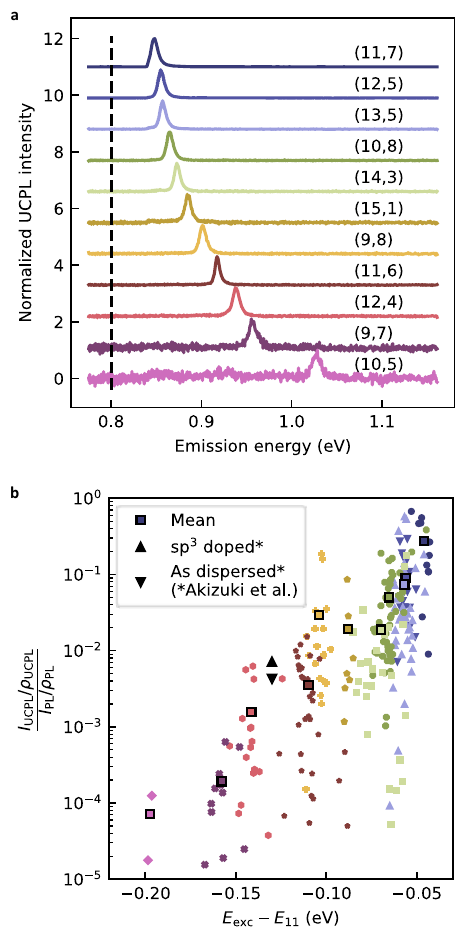}
    \caption{
        \label{fig3}
        (a) Normalized UCPL spectra for various chiralities of SWNTs collected with an excitation energy of 0.800~eV as indicated by the broken line and an excitation power of 1000~$\mu$W.
        (b) The intensity ratio as a function of the energy separation between the excitation and emission energies. The color code of the plot corresponds to the chiralities shown in panel (a). The mean values of the ratio for each chirality along with the ratio for as-dispersed and sp$^3$-doped (6,5) SWNTs in aqueous dispersion~\cite{Akizuki2015a} are plotted together for comparison.
    }
\end{figure}

We now shift our focus to examining UCPL across various chiralities of SWNTs. To assign the chiralities efficiently, we perform three sets of PL measurements with excitation energies of 1.36, 1.46, and 1.60~eV which are near-resonant to many chiralities. Assuming that excitation is close to the $E_{22}$ energy, the chiralities are assigned based on the $E_{11}$ emission energy. UCPL spectra are collected from over 300 individual SWNTs with an excitation energy of 0.800~eV, and representative spectra are shown in Fig.~\ref{fig3}a. We find that UCPL is a universal phenomenon across the chiralities and can be detected up to an emission energy of 0.997~eV, corresponding to an energy gain of 197~meV. Only UCPL spectra with a full width at half maximum of 4--40~meV are analyzed to exclude emission background from substrates. This approach yields 11 chiralities for subsequent analysis of UCPL spectra.

The intensities of UCPL and PL spectra are compared to quantify the dependency of quantum yield on the energy gain relative to the excitation energy~\cite{Akizuki2015a}. We define the intensity ratio as $(I_\mathrm{UCPL}/\rho_\mathrm{UCPL})/(I_\mathrm{PL}/\rho_\mathrm{PL})$ where $I$ is the spectrally integrated intensity and $\rho$ is the excitation power density with subscripts denoting the emission process. The intensities $I_\mathrm{UCPL}$ and $I_\mathrm{PL}$ are obtained by fitting the UCPL and PL spectra, respectively, with a Lorentzian function. The ratios across the chiralities are plotted in Fig.~\ref{fig3}b in which an exponential increase is primarily observed as the energy gain approaches zero. We find that the UCPL intensity is within an order of magnitude compared to the PL intensity
around $E_\mathrm{exc}-E_{11}=-50$~meV where $E_\mathrm{exc}$ is the excitation energy and $E_{11}$ is the emission energy. Assuming that the quantum yield of PL in air-suspended SWNTs with excitation of $E_{22}$ is around 7\%~\cite{Lefebvre2006}, the quantum yield of UCPL is estimated to be about 1\% with an energy gain of approximately 50~meV. It is noteworthy that the intensity ratios for solution-processed nanotubes are on the same order of those for air-suspended SWNTs when comparing at similar energy gains.

We observe a large dispersion in the emission intensity ratio even within the same chiralities (Fig.~\ref{fig3}b). In our PL experiments, we fix the excitation energies at 1.36, 1.46, or 1.60~eV while variations in the $E_{22}$ energies of the order of several~meV occur against a linewidth of 40~meV~\cite{supp,Ishii2015}. Such variations in $E_{22}$ likely contribute to the dispersion of $I_\mathrm{PL}$ influenced by a range of factors. In addition, local strain can be introduced in SWNTs during the growth which reduces the band gap by 100~meV/\%~\cite{Minot2003}. The inhomogeneity of the dielectric environment due to adsorption of water molecules also leads to redshifts~\cite{Homma2013, Uda2018, Uda2018a, Lefebvre2008}. Another factor is the variation in the suspended lengths, altering $I_\mathrm{PL}$ where the conversion from a parity-even dark to parity-odd bright exciton with an efficiency higher than 50\% is possible in micrometer-long air-suspended SWNTs~\cite{Ishii2019}. Furthermore, the difference of the focused beam diameters can amplify the dispersion in the intensity ratio, where the diameter of 2.61~$\mu$m for the UCPL measurements is larger by a factor of $\sim$2 than the diameter for the PL measurements. The larger beam illuminates the entire region of a longer nanotube, resulting in a stronger dependence of the emission intensity on the suspended length.

\begin{figure*}
    \includegraphics{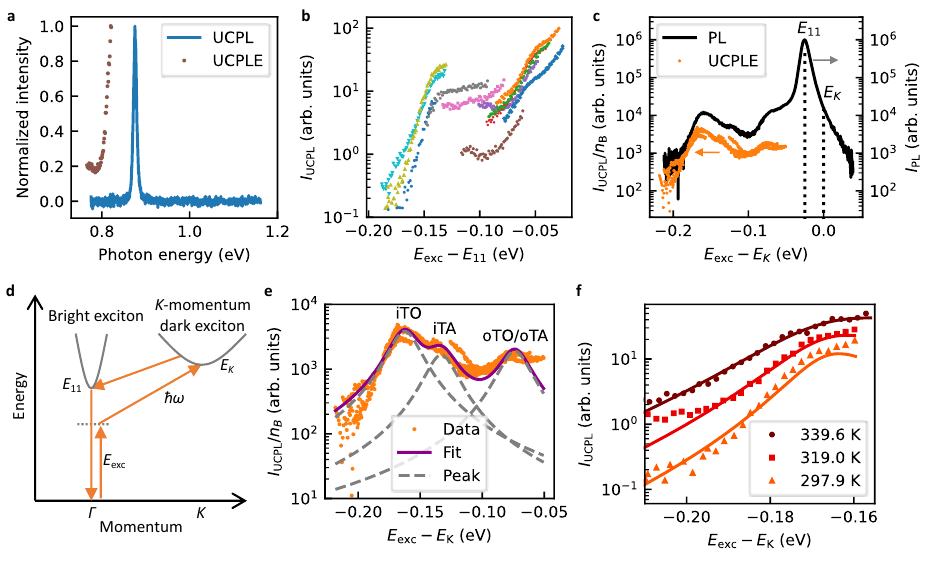}
    \caption{
        \label{fig4}
        (a) UCPL and UCPLE spectra for a (14,3) SWNT with an excitation power of 1000~$\mu$W.
        (b) UCPLE spectra across 11 different chiralities of SWNTs, plotted as a function of the energy gain with an excitation power of 1000~$\mu$W.
        (c) Connected UCPLE spectrum normalized by $n_\mathrm{B}$ plotted on an energy axis relative to $E_K$. A PL spectrum for a (11,3) nanotube is shown for comparison.
        (d) An excitonic band diagram where the arrows indicate transitions.
        (e) The normalized UCPLE spectrum, fitted by a second-order exciton-phonon scattering model. The solid curve is the sum of the fit and the broken lines are the individual components of the fit. $\gamma=12$~meV is used for fits to all the three peaks.
        (f) Temperature-dependent UCPLE spectra for a (9,7) SWNT at an excitation power of 1000~$\mu$W. The solid curves are the fits.
    }
\end{figure*}

\section{Excitation resonances of upconversion photoluminescence}
The large dispersion can be circumvented by examining UCPLE spectra within single SWNTs. In the UCPLE measurements, we employ a frequency-tunable single-mode external cavity laser, sweeping the excitation energy within an available tuning range from 0.756 to 0.838~eV. Although this range of the excitation energy is limited, different ranges of $E_\mathrm{exc}-E_{11}$ can be obtained by consolidating UCPLE spectra with various chiralities. When we sweep the excitation energy, we hold the excitation power constant with a fiber-optic variable optical attenuator.

Figure~\ref{fig4}a shows UCPL and UCPLE spectra for a (14,3) SWNT, in which UCPLE spectrum shows a complex behavior including a slight dip at 0.79~eV and an exponential increase towards higher energy. The entire trend can be captured by compiling UCPLE spectra for the various chiralities (Fig.~\ref{fig4}b; see Fig.~S3 in Supplemental Material for reproducibility~\cite{supp}). In this plot, the UCPL intensities are plotted on a common energy axis relative to the $E_{11}$ emission energy for each chirality. We find that UCPLE intensity exponentially increases up to $-0.14$~eV, exhibits a plateau up to $-0.08$~eV, and increases exponentially with a hump at $-0.06$~eV. This departure from a single-exponential function can be explained by multiple resonances of phonons involved in the upconversion process.

One prominent phonon sideband of interest has been observed in PL, PLE, and absorption spectra, corresponding to the sideband of the $K$-momentum dark singlet exciton coupled with an in-plane transverse optical (iTO) phonon~\cite{Torrens2008,Vora2010,Blackburn2012,Matsunaga2010}. The energy of the $K$-momentum exciton $E_K$ is located 22.2–35.5~meV above the energy of the bright exciton $E_{11}$ for SWNTs with diameters ranging from 0.829 to 0.747~nm~\cite{Vora2010}. In further analysis, we approximate the energy difference $E_K-E_{11}$ as constant at 25~meV across all the chiralities. A PL spectrum on a logarithmic scale for a (11,3) SWNT is displayed on an energy axis relative to $E_K$ (Fig.~\ref{fig4}c; see Fig.~S4 in Supplemental Material for the other chiralities~\cite{supp}). Three sidebands are observed across the chiralities and the energy differences from $E_K$ are determined to be 70, 125, and 159~meV by peak deconvolution (Fig.~S5 in Supplemental Material~\cite{supp}). The third sideband is consistent with the previous reports of the sideband arising from the $K$-momentum iTO phonon mode~\cite{Vora2010,Blackburn2012}. We note that the remaining sidebands have appeared in Refs.~\cite{Kiowski2007,Matsunaga2010,Yasukochi2011,Yoshida2016,Uda2018a}, but their assignments are to be clarified.

To facilitate analysis of the UCPLE spectral features, we seamlessly connect the spectra across the various chiralities by the following procedure. These spectra are initially interpolated on a common energy axis, which allows for a direct comparison of the UCPL intensities. We then minimize the cumulative intensity difference across all permutations of the chiralities, represented by the expression $\sum_{\beta=1}^{N} \sum_{\alpha=1}^{N} \epsilon_{\alpha\beta}$. Here, $N$ is the total number of the chiralities and the subscripts $\alpha$ and $\beta$ are indices representing the chiralities. The intensity difference is calculated as $\epsilon_{\alpha\beta} = \sqrt{(c_\beta I_\beta - c_\alpha I_\alpha)^2}$ where $I$ is the spectrally integrated intensity of the UCPLE spectra. We optimize the fitting coefficients $c_\alpha$ and $c_\beta$ to minimize $\sum_{\beta=1}^{N} \sum_{\alpha=1}^{N} \epsilon_{\alpha\beta}$ under a normalization condition of $c_1 = 1$, obtaining a connected UCPLE spectrum.

In order to compare the sidebands of the PL spectra with the connected UCPLE spectrum, it is needed to take into account the phonon population determining the exciton-phonon scattering rate in the anti-Stokes process. We normalize the intensity of the UCPLE spectrum by the Bose-Einstein distribution of phonons:
\begin{equation}
    \label{bose_einstein}
    n_\mathrm{B}=\frac{1}{\exp \left(\dfrac{\hbar \omega}{k_\mathrm{B} T} \right)-1}
\end{equation}
where $\hbar$ is the Dirac constant, $\omega$ is the angular frequency of phonons, $k_\mathrm{B}$ is the Boltzmann constant, and $T$ is the temperature. In the normalized UCPLE spectrum (Fig.~\ref{fig4}c), three peaks are observed with their positions coinciding with the phonon sidebands in the PL spectra. Notably, the relative intensities of these peaks in the normalized UCPLE spectrum are largely similar to those of the phonon sidebands in the PL spectra. We first interpret the UCPLE peak corresponding to the PL sideband at 159~meV as a peak arising from the well-established sideband of the $K$-momentum exciton coupled with the $K$-momentum iTO phonon mode. The remaining two peaks are likely similar sidebands associated with other $K$-momentum phonon modes.

UCPL involving the $K$-momentum phonons can be understood as a reverse process of the sideband emission observed in PL as depicted in Fig.~\ref{fig4}d. This process begins with the optical excitation of an exciton to an intermediate, virtual state. The process is followed by upconversion into the $K$-momentum dark exciton state, facilitated by coupling with a $K$-momentum phonon with the energy $\hbar \omega$. Rapid scattering then occurs between the $K$-momentum dark and the $\Gamma$-momentum bright states, leading to population equilibrium~\cite{Ishii2019}. The process ends with the recombination of the exciton and emission of a photon while in the bright state.

To quantitatively interpret the spectral features of UCPLE, we develop a theoretical model based on the second-order perturbation theory~\cite{Moura2014}. This model accounts for the interactions between photons and excitons as well as between excitons and phonons. The expression for UCPL intensities is given by
\begin{equation}
    \frac{I(E_\mathrm{exc})}{n_\mathrm{B}} \propto \left|\sum_{j=1}^N \frac{M_j}{E_\mathrm{exc}-\hbar\omega_j-E_K-i\gamma}\right|^2
\end{equation}
where $N$ is the number of the phonon modes, $M_j$ is the matrix element including absorption, exciton-phonon coupling, and emission processes, $\hbar \omega_j$ is the phonon energy for the $j$-th mode, $E_K$ is the energy of the $K$-momentum dark singlet exciton, and $\gamma$ is the damping constant related to the finite lifetime of the intermediate state. We apply $N = 3$ as we observe three peaks in the detection range of the UCPLE (Fig.~4c) and optimize the fit parameters $M_j$, $\omega_j$, and $\gamma$.

The model is fitted to the normalized UCPLE spectrum and shows good agreement except for the region where $E_K-E_{11}$ is close to $-0.05$~eV (Fig.~\ref{fig4}e). The observed discrepancy for smaller $E_K-E_{11}$ is partially due to the direct excitation of the $E_{11}$ state which is not accounted for in our model. The fit to $I(E_\mathrm{exc})/n_\mathrm{B}$ yields phonon energies of $74\pm3$, $133\pm6$, and $163 \pm2$~meV, revealing a close resemblance to the phonon modes at the $K$ point when compared to the phonon dispersion in graphene~\cite{Malard2009,Lazzeri2008,Saito2018}. These UCPLE peaks can be thus explained by the same model where UCPL is a reverse process of the sideband emission from the $K$-momentum dark exciton observed in PL spectra. We are now able to assign the peaks to the out-of-plane transverse optical (oTO)/out-of-plane transverse acoustic (oTA), in-plane transverse acoustic (iTA), and iTO phonon modes at the $K$ point, and the relative amplitudes of the peaks $\propto M_j$ are obtained as $0.298\pm0.006$, $0.278\pm0.010$, and $0.423\pm0.010$, respectively. Notably, $M_j$ extracted from the model indicate that the oTO/oTA and iTA phonon modes exhibit lower scattering rate than the iTO mode. The assignment is further supported by near-linear behavior in the excitation power dependence of UCPL spectra that involves the respective phonon modes (Fig.~S6 in Supplemental Material~\cite{supp}). It should be noted that a signature of the oTO/oTA phonon modes has been detected in PLE spectra~\cite{Vora2010} and predicted by tight-binding calculations~\cite{Perebeinos2005}, which is consistent with our observation. The shoulder at the energy difference of 45~meV relative to $E_{11}$ in the PL spectra~\cite{Kiowski2007,Matsunaga2010,Yasukochi2011,Yoshida2016,Uda2018a} can therefore also be assigned to the oTO/oTA phonon sideband of the $K$-momentum exciton.

We verify our model by examining the temperature dependence of UCPLE spectra. To control the sample temperature during the spectroscopy, the Si substrate is mounted on a flexible resistive foil heater with an insertion of a thermal insulator between the heater and the translation stage. The temperature is monitored with a thermistor integrated into the heater and is ensured to be within an error of 1\% via proportional–integral–derivative feedback. Figure~\ref{fig4}f shows the UCPLE spectra for a (9,7) SWNT at various temperatures. The spectra are fitted using the same model as in Fig.~\ref{fig4}e, where only $\gamma$ is varied while $M_j$ and $\omega_j$ are held constant across the temperatures. The model is able to reproduce the temperature-dependence of the UCPLE spectra that scale with $n_\mathrm{B}$, further reinforcing the validity of our model. We note that this scaling has been observed in the solution-processed SWNTs~\cite{Akizuki2015a}. The values of $\gamma$ are found to be 11, 15, and 21~meV with temperatures of 297.9, 319.0, and 339.6 K, respectively. This increase in $\gamma$ with the temperature corresponds to shortening of the intermediate-state lifetime.

We now discuss the mechanism underpinning the high efficiency of UCPL in our study. The key to this high efficiency lies in the resonant coupling with low-energy phonons, particularly the oTO/oTA modes, which are prevalent at room temperature. Moreover, the phonon-mediated upconversion process benefits from strong exciton-phonon coupling especially near the $K$ points in the Brillouin zone~\cite{Plentz2005, Bindl2011,Vora2010,Torrens2008,Perebeinos2005,Venanzi2023}. The coupling with the $K$-momentum phonons can be stronger when the dimensionality and the excitation energy are lower~\cite{Venanzi2023}. The high efficiency can also be explained by selective population of parity-odd bright excitons in the $E_{11}$ state. For excitation of $E_{22}$ excitons, spontaneous dissociation of the excitons into free electron-hole pairs occurs during the relaxation process~\cite{Kumamoto2014}. The free carriers are then distributed into other singlet and triplet dark exciton states in addition to the bright exciton states~\cite{Gokus2010,Crochet2012}. For excitation below $E_{11}$, excitons at the intermediate state are scattered into the $K$ point, and then reaches the population equilibrium between the $K$-momentum dark and the $\Gamma$-momentum bright states due to the rapid transition process~\cite{Ishii2019}. Considering that the lifetime of the $K$-momentum dark excitons is long and the radiative quantum efficiency of the bright excitons is near unity~\cite{Machiya2022}, almost all excitons in equilibrium should end up emitting light.

\section{Conclusion}
In conclusion, we have investigated the mechanism of UCPL in as-grown SWNTs and shown that the UCPL is a reverse process of the sideband emission from the $K$-momentum dark exciton observed in PL. The comparative analysis of PL and UCPL measurements has confirmed that the emission in the spectra and images originates from the same nanotube. We have attributed the nearly linear excitation power dependence to the one-photon and one-phonon absorption process. The polarization degree of UCPL has been found to be as high as that of PL, which is evidence for the intrinsic nature of the upconversion process. Our investigation has covered eleven distinct SWCNT chiralities exhibiting $E_{11}$ emission in UCPL spectra, which indicates that UCPL is a universal phenomenon across the chiralities. Through UCPLE spectroscopy, we have detected three phonon-related peaks, which arise from the sidebands of the $K$-momentum dark singlet excitons. We have successfully applied a second-order exciton-phonon scattering model, precisely identifying phonon energies and relative matrix element amplitudes for the scattering events with iTO, iTA, and oTO/oTA phonons involved in the upconversion processes. The robustness of our model is further demonstrated by the ability to accurately predict the temperature-dependent UCPLE spectra. Given that the radiative quantum efficiency in SWNTs approaches unity~\cite{Machiya2022}, the upconversion process facilitated by the resonant phonon coupling holds potential for fluorescent cooling with the selective excitation of bright excitons.

\section*{Acknowledgments}
This work is supported in part by JSPS (KAKENHI JP23K23161 to D.K., JP22K14625 to S.F., JP23H00262, JP20H02558 to Y.K.K.), the Precise Measurement Technology Promotion Foundation (PMTP-F to S.F.), and MEXT (ARIM JPMXP1222UT1136). We thank the Advanced Manufacturing Support Team at RIKEN for technical assistance.

\end{document}